\documentclass{article}

\usepackage[english]{babel}
\usepackage{amsmath,comment}
\usepackage{amsthm}
\usepackage{subfigure}
\usepackage{graphicx}
\usepackage{amssymb}

\def\ii{\int\limits_{\mathbb{R}}}
\def\jj{\int\limits_{-h_1}^{\eta}}
\def\jjj{\int\limits_{\eta}^{h_2}}

\title{On the dynamics of internal waves interacting with the equatorial undercurrent}

\author{Alan Compelli $^a$ and Rossen Ivanov $^{b,}$\footnote{Permanent address: School of Mathematical Sciences, Dublin Institute of Technology, Kevin Street, Dublin 8, Ireland; email: rossen.ivanov@dit.ie}\\
\\
$^a$ School of Mathematical Sciences, Dublin Institute of Technology, \\
Kevin Street, Dublin 8, Ireland\\
{\it Email}: alan.compelli@mydit.ie
\\
\\
$^b$ Faculty of Mathematics, Oskar-Morgenstern-Platz 1,\\
 University of Vienna, 1090 Vienna, Austria\\
{\it Email}: rossen.ivanov@univie.ac.at 
}

\begin{document}

\maketitle

\maketitle
\thispagestyle{empty}

\vphantom{\vbox{}}

\begin{abstract}
The interaction of the nonlinear internal waves with a nonuniform current with a specific form, characteristic for the equatorial undercurrent, is studied. The current has no vorticity in the layer, where the internal wave motion takes place. We show that the nonzero vorticity that might be occuring in other layers of the current does not affect the wave motion. The equations of motion are formulated as a Hamiltonian system.
\end{abstract}

{\it Keywords}: Internal waves, Equatorial undercurrent, shear flow, Hamiltonian system.

{\it 2000 Mathematics Subject Classification}: 35Q35, 37K05, 74J30

\section{Introduction}
The dynamics of the Pacific Ocean, within a band of about $2^{\circ}$ latitude from the Equator, is characterised by a pronounced stratification, greater than anywhere else in the ocean; see \cite{FB}. A very sharp layer separates a shallow layer of relatively warm water (and so less dense) near the surface, from a deeper layer of denser, colder water. Since these distinct top and bottom layers differ by density and temperature, the sharp layer between them is termed {\it pycnocline} and sometimes {\it thermocline}. The difference in density across the thermocline is about 1\%. In addition, underlying non-uniform currents occur, see e.g. \cite{MC}. In a band approximately 300 km wide about the Equator, the underlying currents are highly depth-dependent: in a sub-surface layer, typically extending no more than about 100 m down, there is a westward drift that is driven by the prevailing trade winds; just below this there lies the Equatorial Undercurrent (EUC), an eastward jet whose core resides on the thermocline. Below the EUC, the motion dies out rapidly so that, at depths in excess of about 240 m, there is an abyssal layer of essentially still water. Variety of waves in this region is observed, including long waves with wavelengths exceeding 100 km \cite{FB}. There is ample evidence of large-amplitude internal waves, with relatively short wavelengths (typically a few hundreds of metres) and periods about 5-10 min; see \cite{MNS}.

We examine a simplified model of two-dimensional internal geophysical waves propagating along the Equator. Although the Coriolis force plays the role of a waveguide keeping the EUC and the occurring waves to propagate along the equator (and to remain essentially two dimensional) its impact on the dynamics of this two dimensional motion is rather small. Thus, for simplicity we neglect Coriolis forces.

The gravitationally induced internal  waves propagate on the common interface (thermocline/pycnocline) between the upper and lower media of different densities. (We assume for simplicity that both media have constant densities.) In physical reality the ocean surface will have surface waves. We neglect the surface waves and hence consider the top surface to be a flat boundary like a rigid lid. For long and intermediate surface waves coupled with internal waves such an approximation is justified, cf. \cite{FB,CJ}.
The boundary underneath is by an impermeable flat bed, which is a relevant approximation for the ocean bed, in the equatorial Pacific. 

Recently, explicit nonlinear solutions for equatorially trapped waves were obtained in the Lagrangian framework, see \cite{C1,H}. Correspondingly, solutions for the internal waves have been described in \cite{C2}. However, both these types of solution are restricted to relatively short wavelengths, and they are realistic representations of the flow only for the region near the surface and in the neighbourhood of the thermocline, respectively.   
   
By examining the governing equations of the system we provide a Hamiltonian formulation of the equations and with a variable transformation we show that it has a canonical Hamiltonian structure. The Hamiltonian formulation of surface waves over deep water has been noticed by Zakharov \cite{Za} and since then the Hamiltonian models have been used extensively for various approximations with linear and nonlinear PDEs, e.g. \cite{Craig1,Craig2,CIP,CI}.


As per Figure 1 we define the lower medium $\Omega_1$ as the domain $\{(x, y)\in\mathbb{R}^2: -h_1< y < \eta(x,t)\}$, the upper medium $\Omega_2$ as the domain $\{(x, y)\in\mathbb{R}^2: \eta(x,t)< y < h_2\}$ and the entire system $\Omega_{1,2}$ as the domain $\{(x, y)\in\mathbb{R}^2:-h_1< y < h_2\}$ where $\{y=\eta(x,t)\}$ describes the elevation of the common interface. The subscript $c$ will be used to denote evaluation at the common interface.
The shear flow in a general form is 
\begin{equation}
\label{phi_def}
  U(y)=      \left\lbrace
        \begin{array}{lcl}
        - \sigma, \qquad y=h_2
        \\
        \kappa,\qquad l_2\ge y \ge - l_1
        \\
        0, \qquad   y = -h_1
        \end{array}
        \right.
\end{equation}

\noindent and moreover $U(y) $ is a continuous function for $h_2 \ge y \ge l_2$ and $-l_1 \ge y \ge -h_1$. It is apparent that $h_2>l_2>0> -l_1>-h_1$.\\
The constants $\sigma,$ $\kappa$, $l_{1,2}$,  $h_{1,2} $ are positive (see Fig. 1). We use the subscript notation $i=\{1,2\}$ to represent the lower and upper media respectively. The next assumption is that the pycnocline is always in the strip $l_2\ge y \ge - l_1$, i.e.  $$l_2\ge \eta(x,t) \ge - l_1$$  for all $x,t$ so that all internal waves take place in this {\it strip}.

The velocities in the two media are decomposed into 'wave' and 'current' parts:
\begin{equation}
\label{phi_def}
        \left\lbrace
        \begin{array}{lcl}
        u_i  =  {\varphi}_{i,x} +U_i(y) =\psi_{i,y}
        \\
        v_i =  {\varphi}_{i,y}=- \psi_{i,x}
        \end{array}
        \right.
\end{equation}

\noindent where without any ambiguity
\begin{equation}
\label{U12}
        \left\lbrace
        \begin{array}{lcl}
       U_1(y)=U(y), \qquad h_2 \ge y \ge \eta(x,t)
        \\
      U_2(y)=U(y), \qquad \eta(x,t)\ge y \ge -h_1. 
        \end{array}
        \right.
\end{equation}

The non-lateral velocity flow, with propagation in the positive $x$-direction, is given by ${\bf{V}}_i(x,y,z)=(u_i,v_i,0)$ 
$\rho_1$ and $\rho_2$ are the respective constant densities of the lower and upper media and stability is given by the condition that
\begin{alignat}{2}
\label{stability}
\rho_1>\rho_2.
\end{alignat}

\begin{figure}
\begin{center}
\includegraphics[width=.7\textwidth]{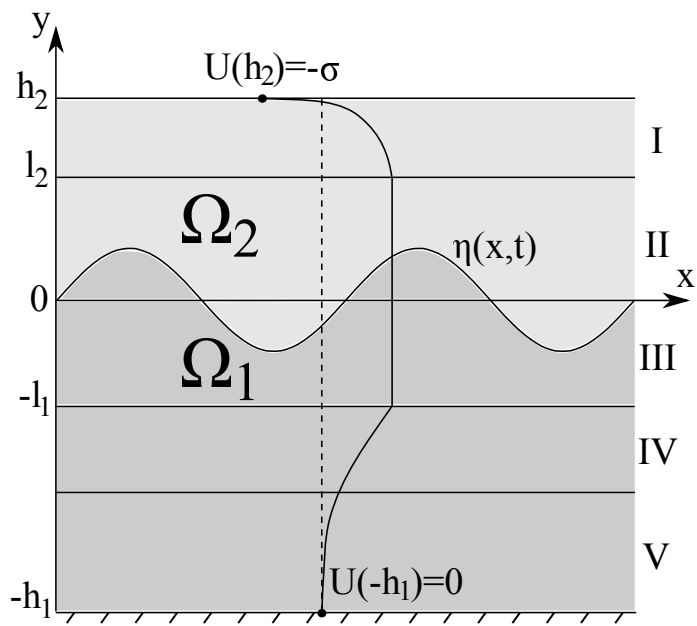}
\end{center}
\caption{Setup: two domains $\Omega_1$ and $\Omega_2$ with densities $\rho_1$ and $\rho_2$.  The shear profile $U(y)$ is provided qualitatively. The five layers I -- V correspond to the description in \cite{CJ}. The important feature is that the wave motion takes place in layers II and III where the vorticity is zero and $U(y)$ is constant. In the other layers $U(y)$ is continuous.}
\end{figure}

We assume that for large $|x|$ the amplitude of $\eta$ attenuates and hence make the following assumptions
\begin{equation}
\label{Assump1}
\lim_{|x|\rightarrow \infty}\eta(x,t)=0, \qquad
\lim_{|x|\rightarrow \infty}\varphi_i(x,y,t)=0.
\end{equation}

\section{Governing Equations}
We write Euler's equation as:
\begin{alignat}{2}
{\bf{V}}_{i,t}+({\bf{V}}_i.\nabla){\bf{V}}_i=-\frac{1}{\rho_i}\nabla P_i+{\bf{g}}+{\bf{F}}_c
\end{alignat}
where $P_i=\rho_i g y+p_{\mbox{atm}}+p_i$ is the pressure at a depth $y$, $p_{\mbox{atm}}$ is (constant) atmospheric pressure, $p_i$ is the dynamic pressure due to the wave motion, $g$ is the acceleration due to gravity (where $y$ points in the opposite direction to the center of gravity), ${\bf{g}}$ is the force due to gravity per unit mass, and
the Coriolis force is neglected for simplicity. Its effect is not significant and also it is important to keep the waves one-dimensional.

Applying Equations (\ref{phi_def}) and noticing that in the strip $l_2 \ge \eta(x,t) \ge -l_1$  we have $U(y)=\kappa=\mathrm{const.}$ and velocities
\begin{equation}
\label{strip_vel}
        \left\lbrace
        \begin{array}{lcl}
        u_i  =  {\varphi}_{i,x} +\kappa = \psi_{i,y}
        \\
        v_i =  {\varphi}_{i,y}=- \psi_{i,x}
        \end{array}
        \right.
\end{equation}
and thus there is no vorticity in the strip, i.e. $\Delta \psi(x,y)=0$. The equations can be written as
\begin{alignat}{2}
\nabla\Big(\varphi_{i,t}+\frac{1}{2}(\nabla \psi_i)^2 \Big)=\nabla\Big(-g y-\frac{p_i}{\rho_i}\Big)
\end{alignat}
where $\nabla=(\partial_x,\partial_y)$.
At the interface $p_1=p_2=p_c$ therefore we have the conservation law (Bernoulli condition) at the interface
\begin{equation}
\label{Bernoulli}
\rho_1\Big((\varphi_{1,t})_c+\frac{1}{2}(\nabla \psi_1)_c^2 +g\eta\Big)=\rho_2\Big((\varphi_{2,t})_c+\frac{1}{2}(\nabla \psi_2)_c^2 +g\eta\Big).
\end{equation}
Writing the Bernoulli condition in terms of 'wave' potentials $\varphi_i$ only
\begin{equation}
\label{Bernoulli2}
\big ({(\rho_1\varphi_1}-\rho_2 \varphi_2)_t+ \kappa {(\rho_1\varphi_1}-\rho_2 \varphi_2)_x\big )_c+\rho_1\frac{1}{2}(\nabla \varphi_1)_c^2 -\rho_2 \frac{1}{2}(\nabla \varphi_2)_c^2+g\eta(\rho_1 - \rho_2)=0 .
\end{equation}
The difference with the case without current is only the shift
$$ \partial_t \rightarrow \partial_t + \kappa \partial_x.$$
This also suggests the definition of $\xi_i:=(\varphi_i)_c=\varphi_i(x,\, \eta(x,\,t),\,t)$  as the interface velocity potential and hence the definition of  {\footnote{One has to be very careful since $\xi_t =( (\rho_1\varphi_1-\rho_2 \varphi_2)_c)_t\neq ( (\rho_1\varphi_1-\rho_2 \varphi_2)_t)_c$}} 
\begin{alignat}{2}
\xi:=\rho_1\xi_1-\rho_2\xi_2.
\end{alignat}
We will also use the following kinematic boundary conditions:
\begin{equation}
\label{KBC}
        \left\lbrace
        \begin{array}{lcl}
        \eta_t = -\eta_x\big((\varphi_{i,x})_c+ \kappa\big) + (\varphi_{i,y})_c
        \\
        (\varphi_{1,y})_b=(\varphi_{2,y})_l=0
        \end{array}
        \right.
\end{equation}
noting that ${\bf{V_1}}(x,-h_1,0)=(u_1,0,0)$ and ${\bf{V_2}}(x,h_2,0)=(u_2,0,0)$, where the subscripts $b$ and $l$ denote evaluation at the bottom (lower boundary) and lid (upper boundary) respectively. Again we observe only a shift $ \partial_t \rightarrow \partial_t + \kappa \partial_x $ in comparison with the case $U(y) \equiv 0$:
$$   \eta_t + \kappa \eta_x = -(\varphi_{i,x})_c  \eta_x + (\varphi_{i,y})_c  $$
Therefore the time evolution of the variables, describing the dynamics of the internal waves $\xi, \eta $ does not depend on the shear outside the 'strip'; the constant flow in the strip leads only to a velocity shift with the velocity of the current $\kappa$ (in comparison to the currents free case).

\section{Hamiltonian Formulation}
We will analyse the phenomenon from the Hamiltonian point of view. If we consider the system under study as an irrotational system the Hamiltonian, $H$, is given by the sum of the kinetic and potential energies as:

\begin{equation} 
H= \frac{1}{2}\ii  \jj\rho_1 (u_1^2 +v_1^2 )\, dy\,dx +\frac{1}{2}\ii  \jjj \rho_2(u_2^2 +v_2^2 ) \, dy\,dx +\frac{1}{2}\ii(\rho_1-\rho_2)g\eta^2 \, dx.
\end{equation}

\noindent We try to express the Hamiltonian in terms of $\xi, \eta$. The kinetic energy of the lower layer is
\begin{equation} \nonumber
\begin{split}
K_1= &\frac{1}{2}\ii  \jj\rho_1 \Big[( \varphi_{1,x} +U_1(y))^2 + (\varphi_{1,y})^2 \Big] \, dy\,dx \\=&
\frac{\rho_1}{2}\ii  \jj (\nabla \varphi_1)^2 \, dy\,dx +  \rho_1 \ii  \jj( \varphi_{1,x} )U_1(y)  \, dy\,dx + \frac{\rho_1}{2}\ii  \jj U_1^2(y) \, dy\,dx.
\end{split}
\end{equation}
We evaluate
\begin{eqnarray}
&\phantom{o}&\ii  \jj( \varphi_{1,x} )U_1(y)  \, dy\,dx = \int_{{\mathbb{R}}}\left(\int_{-h_1}^{-l_1}+ \int_{-l_1}^{\eta(x,t)}\right)( \varphi_{1,x} )U_1(y)  \, dy\,dx \nonumber \\
&=&\int_{-h_1}^{-l_1}\Big(\int_{{\mathbb{R}}}( \varphi_{1,x} (x,y) ) \, dx \Big) U_1(y)  \, dy  +\int_{{\mathbb{R}}} \int_{-l_1}^{\eta(x,t)}( \varphi_{1,x} ) \kappa  \, dy\,dx \nonumber \\
&=&\int_{-h_1}^{-l_1}\Big(\int_{{\mathbb{R}}}( \varphi_{1,x} (x,y) ) \, dx \Big) U_1(y)  \, dy  +\kappa\int_{{\mathbb{R}}} \left(-\varphi_1(x,\eta) \eta_x + \partial_x\int_{-l_1}^{\eta(x,t)} \varphi_1   \, dy \right)\,dx \nonumber \\
&=&-\kappa \int_{{\mathbb{R}}}\xi_1 \eta_x \, dx
\end{eqnarray}

\noindent since the integrals over total $x$ derivatives are zeroes due to the boundary conditions \eqref{Assump1}. The integral
\begin{eqnarray}
\ii  \jj U_1^2(y) \, dy\,dx=\int_{{\mathbb{R}}}\left(\int_{-h_1}^{-l_1}+ \int_{-l_1}^{\eta(x,t)}\right) U_1^2(y) \, dy\,dx  \nonumber\\
=\int_{{\mathbb{R}}}\left(\int_{-h_1}^{-l_1} U_1^2(y) \, dy \right) \,dx+\int_{{\mathbb{R}}} \left(\int_{-l_1}^{\eta(x,t)} \kappa^2 \, dy \right)\,dx  \nonumber\\
=\int_{{\mathbb{R}}}\left(\int_{-h_1}^{-l_1} U_1^2(y) \, dy \right) \,dx+\kappa^2\int_{{\mathbb{R}}} \left(\eta(x,t) + l_1  \right)\,dx 
\end{eqnarray}

\noindent gives an irrelevant constant, which can be absorbed by a proper renormalization and does not produce any dynamic terms, since the average elevation of the thermocline by definition is zero at $y=0$, i.e.  $\int_{{\mathbb{R}}} \eta(x,t) \,dx =0.$ The integrals over the upper layer produce

$$\rho_2 \kappa \int_{{\mathbb{R}}}\xi_2 \eta_x \, dx  $$ and altogether the kinetic terms produce the extra term

$$ \kappa \int_{{\mathbb{R}}} (\rho_2\xi_2 - \rho_1 \xi_1)\eta_x \, dx  = -\kappa \int_{{\mathbb{R}}} \xi \eta_x \, dx.$$

\noindent This quantity itself is a conservation law, related to the energy of the overall translation of the system by a velocity $\kappa$, see \cite{Craig2}. The Hamiltonian of the system is given by:
\begin{equation} 
\begin{split}
H= & \frac{1}{2}\ii  \jj\rho_1 (\nabla \varphi_1)^2 \, dy\,dx+ \frac{1}{2}\ii  \jjj \rho_2(\nabla \varphi_2)^2 \, dy\,dx \\
+&\frac{1}{2}\ii(\rho_1-\rho_2)g\eta^2 \, dx -\kappa \int_{{\mathbb{R}}} \xi \eta_x \, dx.
\end{split}
\end{equation}

We notice that the Hamiltonian does not depend on the $U(y)$ outside the {\it strip}. In order to express it in terms of the desired variables we introduce the Dirichlet-Neumann operators $G_i(\eta)$ given by
\begin{alignat}{2}
G_i(\eta)\xi_i=(\varphi_{i,{\bf{n}}_i})\sqrt{1+(\eta_x)^2},
\end{alignat}
where $\varphi_{i,{\bf{n}}_i}$ is the normal derivative of the velocity potential $\varphi_i$, at the surface, for an outward normal ${{\bf{n}}_i}$, see more details in \cite{Craig1,Craig2}. We define
\begin{alignat}{2}
B:=\rho_1 G_2(\eta)+\rho_2 G_1(\eta).
\end{alignat}

We can express the Hamiltonian in terms of the conjugate variables $(\eta,\xi)$ as
\begin{multline}
\label{Hametaxi}
H(\eta,\xi)= \frac{1}{2}\ii  \xi \big(G_1(\eta) B^{-1}G_2(\eta)\big)\xi \,dx+ \frac{1}{2}\ii(\rho_1-\rho_2)g\eta^2 \, dx -\kappa \int_{{\mathbb{R}}} \xi \eta_x \, dx\\
=H_0(\eta,\xi)-\kappa \int_{{\mathbb{R}}} \xi \eta_x \, dx. 
\end{multline}
where $H_0(\eta,\xi)$ is the Hamiltonian in absence of shear, $U(y)\equiv 0$ everywhere. 

From (\ref{KBC}), $\eta_t=\big(\psi_{i}(x,\eta,t)\big)_x$ and \eqref{Bernoulli2} 
can be written as a canonical Hamiltonian system
\begin{equation}
\label{EOMsys}
        \left\lbrace
        \begin{array}{lcl}
        \eta_t=\delta_{\xi} H ,
        \\
        \xi_t=-\delta_{\eta} H.
        \end{array}
        \right.
\end{equation}
\noindent or
\begin{equation}
\label{Hsys}
        \left\lbrace
        \begin{array}{lcl}
        \eta_t=\delta_{\xi} H_0 -\kappa \eta_x,
        \\
        \xi_t=-\delta_{\eta}H_0-\kappa \xi_x.
        \end{array}
        \right.
\end{equation}

\noindent For the shifted variables $X=x-\kappa t$ and $T=t$, $\partial_T=\partial_t+\kappa \partial_x $
\begin{equation}
\label{Hsys}
        \left\lbrace
        \begin{array}{lcl}
        \eta_T=\delta_{\xi} H_0 ,
        \\
        \xi_T=-\delta_{\eta}H_0 .
        \end{array}
        \right.
\end{equation}

 Therefore this system of equations has the usual canonical Hamiltonian form (same as the Hamiltonian form when there is no current) after the Galilean change of the coordinates, i.e. in a system travelling with velocity $\kappa$ together with the flow in the strip  $l_2 \ge y \ge -l_1$.

\section{Linearization of the model equations}
By Taylor expanding the Dirichlet-Neumann operator we can represent it in terms of orders of $\eta$ as
\begin{alignat}{2}
\label{G_expanded}
G_i(\eta)=\sum_{j=0}^{\infty} G_{ij}(\eta)
\end{alignat}
with the constant, linear and quadratic terms given as
\begin{alignat}{2}
\nonumber
G_{i0}&=D\tanh(h_iD)\nonumber \\
G_{11}(\eta)&=D \eta  D -G_{10} \eta G_{10}\nonumber \\
G_{21}(\eta)&=-D \eta  D +G_{20} \eta G_{20}\nonumber \\
G_{i2}(\eta)&=-\frac{1}{2}\left(D^{2} \eta^2 G_{i0} -2 G_{i0} \eta G_{i0} \eta G_{i0}+G_{i0}\eta^2 D^2  \right) \nonumber
\end{alignat}
where the operator $D$ is a Fourier multiplier equivalent to both the operation $-i\partial_x$ and the wavenumber $k$, i.e.
\begin{alignat}{2}
\label{Fourier_M}
D=-i\partial_x=k.
\end{alignat}

The operator $B$ can therefore be expressed as
\begin{alignat}{2}
\label{B_expanded}
B=\rho_1\sum_{j=0}^{\infty} G_{2j}(\eta)+\rho_2\sum_{j=0}^{\infty} G_{1j}(\eta),
\end{alignat}
and, also, the Hamiltonian can be represented as
\begin{alignat}{2}
\label{H_expandj_mu1}
H(\eta,\xi)=\sum_{j=0}^{\infty}\,H^{(j)}(\eta,\xi).
\end{alignat}

Using explicit expressions  $H^{(2)}$ in terms of $\eta$ and $\xi$ is given by
\begin{multline}
\label{lin_Ham}
H^{(2)}(\eta,\xi)
= \frac{1}{2}\ii  \xi\frac{D\tanh(h_1 D)\tanh(h_2 D)}{\rho_1 \tanh(h_2 D)+\rho_2 \tanh(h_1 D)}\xi\,dx\\
+\frac{1}{2}\ii(\rho_1-\rho_2)g{\eta}^2 \, dx-\kappa \int_{{\mathbb{R}}} \xi \eta_x \, dx.
\end{multline}

Recalling (\ref{Hsys}) we calculate the linearized equations of motion as
\begin{alignat}{2}
\label{Lineta_t}
\eta_t + \kappa \eta_x&=  \frac{D\tanh(h_1 D)\tanh(h_2 D)}{\rho_1 \tanh(h_2 D)+\rho_2 \tanh(h_1 D)}\xi,\\
\label{Linxi_t}
\xi_t + \kappa \xi_x&=-(\rho_1-\rho_2)g \eta.
\end{alignat}

Looking for solutions in the form 
\begin{equation}
        \left\lbrace
        \begin{array}{lcl}
        \eta(x,t)=\eta_0e^{i(kx-\Omega(k) t)}
        \\
        \xi(x,t)=\xi_0e^{i(kx-\Omega(k) t)}
        \end{array}
        \right.
\end{equation}
where, as usual $k$ is the wave number, $\Omega(k)$ is the angular frequency and $c(k)=\Omega/k$ is the wave speed, from the compatibility of the two equations we obtain the dispersion law

\begin{alignat}{2} \label{dl}
 (c-\kappa)^2=\frac{g(\rho_1-\rho_2)\tanh(h_1 k)\tanh(h_2 k)}{k\left(\rho_1 \tanh(h_2 k)+\rho_2 \tanh(h_1 k)\right)} .
\end{alignat}
In the case of infinite media as $h_i\rightarrow\infty$ then $\tanh({h_i})\rightarrow1$ and \eqref{dl} becomes
\begin{alignat}{2}
 ( c- \kappa)^2= \frac{g}{k}\frac{(\rho_1-\rho_2)}{(\rho_1+\rho_2)}.
\end{alignat}

The long-wave approximation of \eqref{dl} is the case $k\to 0$ which gives

\begin{alignat}{2}
 (c-\kappa)^2=\frac{g h_1  h_2(\rho_1-\rho_2) }{\rho_1 h_2 +\rho_2 h_1 } .
\end{alignat}

 Lastly, the consideration of the system as a single media system, i.e. $\rho_2\rightarrow 0$ in \eqref{dl} gives
\begin{alignat}{2}
 (c- \kappa)^2=\frac{g}{k}\tanh(h_1 k),
\end{alignat}
which is the well known dispersion relation for the linear approximation of gravity water waves in a single medium irrotational system \cite{C}.

\section{Conclusions}

The shear flow in the layers, away from the thermocline (the strip where the internal waves occur) does not affect the dynamics of the internal waves.
The results are complementary to the case where the vorticity is constant but with a finite jump at $y=\eta(x,t)$ considered in \cite{AC1, AC2}. In the latter case vorticity does affect the wavespeed of the internal waves, cf. with \cite{CI}.

\section*{Acknowledgements}

A.C. is funded by the Fiosraigh Scholarship Programme of Dublin Institute of Technology.  The support of the FWF Project I544-N13 ``Lagrangian kinematics of water waves'' of the Austrian Science Fund is gratefully acknowledged by R.I. The authors are grateful to Prof. A. Constantin for many valuable discussions and to an anonymous referee for several very useful suggestions.

\end{document}